\documentclass[preprint,prc,aps,tightenlines,floats,showpacs,showkeys,nofootinbib]{revtex4-1}
\usepackage{amsmath}
\usepackage{amssymb}
\usepackage[dvips]{graphicx,color}
\usepackage{dcolumn}
\usepackage{epsfig}
\usepackage{bm}

\usepackage{color}

\begin{document}

\title{Impact of $\pi N \to \pi \pi N$ data on determining high-mass nucleon resonances}

\author{Hiroyuki Kamano}
\affiliation{Research Center for Nuclear Physics, Osaka University, Ibaraki, Osaka 567-0047, Japan}

\begin{abstract}
Motivated by an experimental proposal for the measurement of the $\pi N \to \pi \pi N$ 
reactions at J-PARC, we examine the potential impact of the $\pi N \to \pi \pi N$ 
cross section data on the determination of the resonance parameters 
of the high-mass $N^\ast$ states.
For this purpose, we make use of the ANL-Osaka dynamical coupled-channels model, 
which has been developed recently through a combined analysis of
the unpolarized cross section as well as polarized observables from
pion- and photon-induced $\pi N$, $\eta N$, $K\Lambda$, and $K \Sigma$ 
production reactions off a proton target.
We present predictions for the $\pi N \to \pi \pi N$ total cross sections
and invariant mass distributions,
and demonstrate that the $\pi N \to \pi \pi N$ differential cross section data
can indeed be a crucial source of information for understanding 
$N^\ast \to \pi\Delta, \rho N, \sigma N\to \pi\pi N$ decay
of the high-mass $N^\ast$ states.
\end{abstract}
\pacs{14.20.Gk, 13.75.Gx, 13.60.Le}

\maketitle

\section{Introduction}
\label{sec:intro}

Most of the high-mass nucleon resonances ($N^\ast$),
of which complex pole mass $M_R$ satisfies ${\rm Re}(M_R)\gtrsim 1.6$ GeV,
are known to decay strongly to the three-body $\pi \pi N$ continuum states.
Furthermore, it is also known that the double-pion production processes dominate 
the cross sections of $\pi N$ and $\gamma N$ reactions above $W\sim 1.6$ GeV.
These facts naturally lead to the realization
that the reaction analyses including double-pion production data
are indispensable to establishing the mass spectrum
of the high-mass $N^\ast$ states, which remains poorly understood 
despite years of study.

Such analyses, however, are not easy to pursue at the present time mainly due to 
the lack of the data of $\pi N \to \pi \pi N$
in the relevant energy region above $W\sim 1.6$ GeV.
In fact, in this energy region practically no
differential cross section data of $\pi N \to \pi \pi N$ 
are available for detailed partial wave analyses.
(See, e.g., Refs.~\cite{manley,kjlms09,add-exp1,add-exp2} for 
the details of the current situation of the world data of $\pi N \to \pi \pi N$.)
Although the high statistics data of $\gamma N \to \pi \pi N$ are becoming available 
from electron/photon beam facilities such as JLab, Bonn, Mainz, SPring-8, and ELPH
at Tohoku University,
the $\pi N \to \pi \pi N$ data are still highly desirable 
because it is free from electromagnetic interactions
of hadrons that bring additional complications to the analyses.

It is therefore quite encouraging to see the approval of an
experimental proposal at J-PARC to develop plans for performing 
precise measurements of $\pi^\pm p \to \pi \pi N$ 
above $W\sim 1.6$ GeV (J-PARC E45~\cite{hicks-prop}).
Once such precise data are available and included in partial wave analyses, 
the current $N^\ast$ mass spectrum might require significant modifications. 
Besides this, understanding of the final state interactions of the 
three-body $\pi \pi N$ state is also very important for constructing
neutrino-induced reaction models in the GeV-energy region.
A precise knowledge of neutrino-nucleon/nucleus interactions
is required for the determination of leptonic $CP$ phase and the neutrino mass hierarchy 
through accelerator and atmospheric neutrino experiments 
(see, e.g., Refs.~\cite{neutrino,neuint}).

In this work, we examine whether the $\pi N \to \pi \pi N$ data 
can provide crucial constraints on the determination of 
the $N^\ast$ resonance parameters such as pole positions and decay branching ratios.
For this purpose, we make use of a reaction model recently described in Ref.~\cite{dcc8},
which is based on the ANL-Osaka dynamical coupled-channels (DCC) approach~\cite{msl}.
In this approach, the amplitudes of meson production reactions off a nucleon are
given by solving the unitary coupled-channels integral equations.
As a result, it is ensured that the amplitudes satisfy
the two- and three-body unitarity.
In the DCC model of Ref.~\cite{dcc8}, 
the $\pi N$, $\eta N$, $\pi \pi N(\pi\Delta,\rho N,\sigma N)$,
$K\Lambda$, and $K\Sigma$ channels are taken into account
and the model parameters are determined by 
a comprehensive analysis of both pion- and photon-induced 
$\pi N$, $\eta N$, $K\Lambda$, and $K\Sigma$ production reactions off a proton target,
where the data up to $W=2.3$ GeV are taken into account for $\pi N \to \pi N$ 
and those up to $W=2.1$ GeV are for the other reactions.

\section{ANL-Osaka DCC model for the $\pi N \to \pi \pi N$ reaction}
\label{sec:ao}

\begin{figure}
\includegraphics[width=0.8\textwidth,clip]{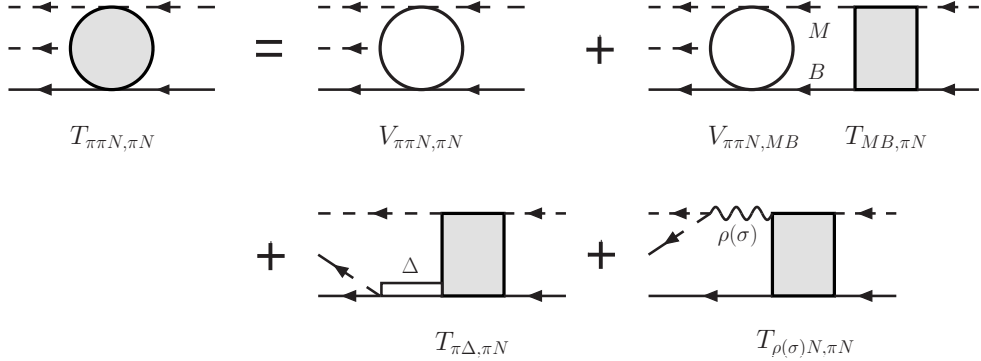}
\caption{
Graphical representation of Eq.~(\ref{eq:amp}).
\label{fig:amp}
}
\end{figure}

In the ANL-Osaka DCC approach~\cite{kjlms09,msl}, the $\pi N\to\pi \pi N$ amplitude 
has the graphical representation shown in Fig.~\ref{fig:amp} and 
is written explicitly as,
\begin{equation}
T_{\pi \pi N,\pi N} 
=
T^{\text{dir}}_{\pi \pi N, \pi N} 
+\sum_{MB=\pi\Delta,\sigma N, \rho N}T^{MB}_{\pi \pi N,\pi N} \,,
\label{eq:amp}
\end{equation}
with
\begin{eqnarray}
T^{\text{dir}}_{\pi \pi N, \pi N}
&=&
 V_{\pi \pi N, \pi N}+
\sum_{MB} 
V_{\pi\pi N,MB} G_{MB} T_{MB,\pi N} \,,
\label{eq:tdir}
\\
T^{\pi\Delta}_{\pi\pi N,\pi N}
&=&
\Gamma_{\pi N, \Delta} G_{\pi \Delta} T_{\pi\Delta, \pi N} \,,
\\
T^{\rho N}_{\pi\pi N,\pi N}
&=&
\Gamma_{\pi \pi, \rho} G_{\rho N} T_{\rho N, \pi N} \,,
\\
T^{\sigma N}_{\pi\pi N,\pi N}
&=&
\Gamma_{\pi \pi, \sigma} G_{\sigma N} T_{\sigma N, \pi N} \,.
\end{eqnarray}
Here $V_{\pi \pi N, MB}$ is a potential describing 
the direct two-body to three-body transition processes~\cite{kjlms09};
$G_{MB}$ is the Green's function of the $MB$ channel;
$\Gamma_{\pi N,\Delta}$, $\Gamma_{\pi\pi,\rho}$, and $\Gamma_{\pi\pi,\sigma}$ 
are the decay vertices for $\Delta\to\pi N$, 
$\rho \to \pi \pi$, and $\sigma \to \pi \pi$, respectively.
The summation $\sum_{MB}$ in Eq.~(\ref{eq:tdir}) runs over 
$MB=\pi N, \eta N, \pi \Delta, \rho N, \sigma N, K\Lambda, K\Sigma$.
As for the two-body amplitudes $T_{MB, \pi N}$,
we employ those obtained in our recent DCC analysis~\cite{dcc8},
which is completely unitary in the $\pi N$, $\eta N $, $\pi\pi N(\pi\Delta,\rho N,\sigma N)$, 
$K\Lambda$, and $K\Sigma$ channel space.
The detailed description of the two-body amplitudes,
meson-baryon Green's functions, and decay vertices in the above equations 
can be found in Ref.~\cite{dcc8} and will not be presented here.

\section{Predicted total cross sections of the $\pi N \to \pi \pi N$ reaction}
\label{sec:predict}

\begin{figure}
\includegraphics[width=0.8\textwidth,clip]{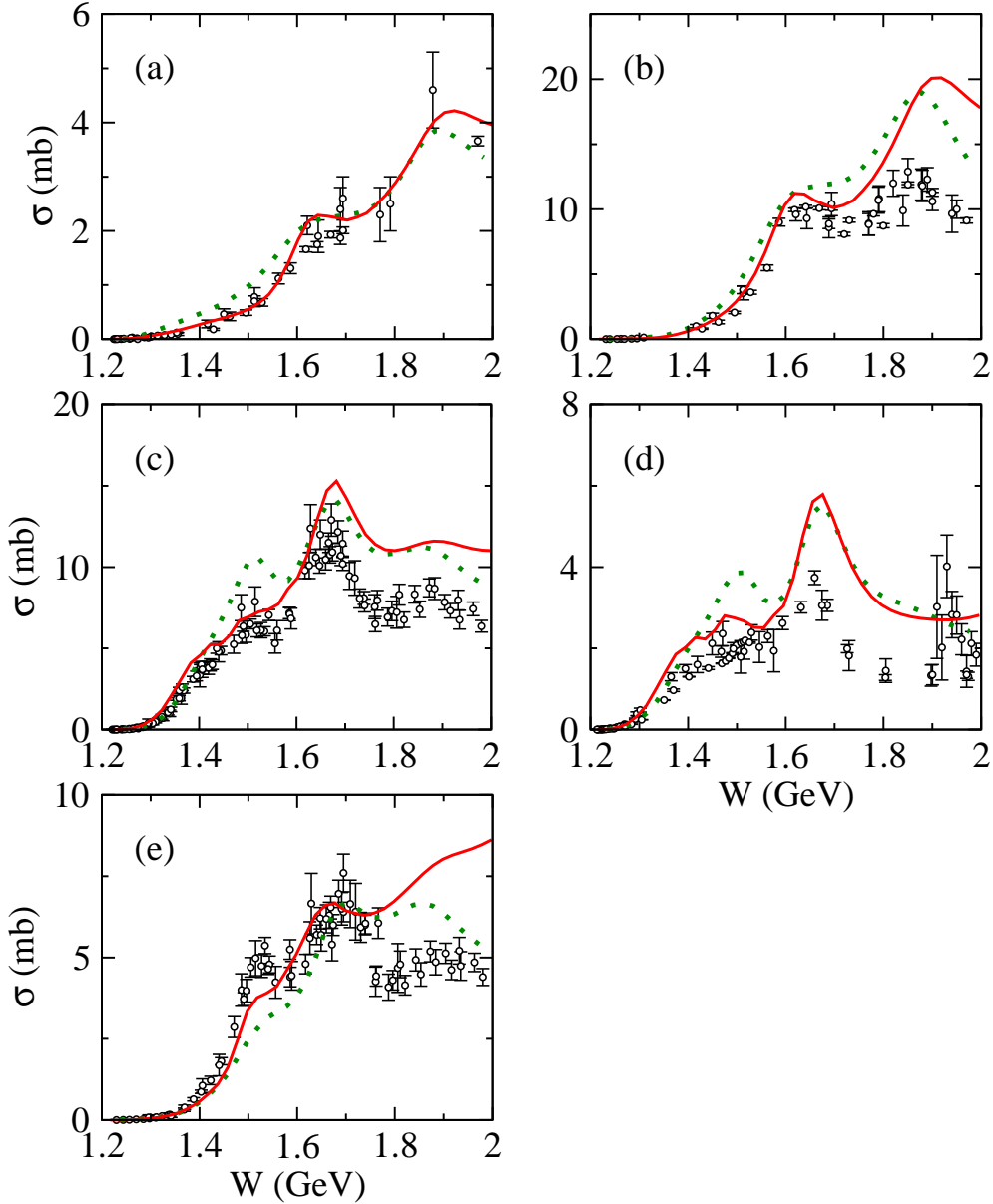}
\caption{
(Color online)
The $\pi N\to \pi\pi N$ total cross sections predicted with the ANL-Osaka DCC models. 
The red solid curves are from our new model recently developed in Ref.~\cite{dcc8},
while the green dotted curves are from our early model~\cite{kjlms09}.
The reaction channels are for
(a)~$\pi^+ p \to \pi^+ \pi^+ n$,
(b)~$\pi^+ p \to \pi^+ \pi^0 p$,
(c)~$\pi^- p \to \pi^+ \pi^- n$,
(d)~$\pi^- p \to \pi^0 \pi^0 n$, and
(e)~$\pi^- p \to \pi^- \pi^0 p$.
See Refs.~\cite{manley,kjlms09} and references therein for the data.
\label{fig:tcs}
}
\end{figure}
Figure~\ref{fig:tcs} shows the $\pi N \to \pi \pi N$ total cross sections up to $W=2$~GeV. 
The red solid curves are the prediction of our new DCC model developed in Ref.~\cite{dcc8}.
As a comparison, we also present the prediction of our early DCC model~\cite{kjlms09,jlms07}
(the green dotted curves),
in which the amplitudes are unitary in the $\pi N$, $\eta N$, and $\pi\pi N(\pi\Delta,\sigma N,\rho N)$ 
channel space and are determined by analyzing the $\pi N \to\pi N$ scattering up to $W=2$ GeV.
Even without any adjustment of the model parameters to the $\pi N \to \pi \pi N$ data,
our models reproduce the available total-cross-section data to better than 
$\sim 20 \%$ ($\sim 60 \%$) accuracy at $W < 1.6$ GeV ($W > 1.6$ GeV).
As can be seen in Fig.~\ref{fig:tcs}, the new DCC model has a better description 
of the total cross sections below $W = 1.6$~GeV.
It is noted that the multichannel unitarity including the three-body 
$\pi \pi N$ channel, which gives a significant constraint on 
the transition probabilities between the reaction channels maintained
in our model, makes it possible to have reasonable predictions for 
the $\pi N \to \pi \pi N$ observables
in this wide energy range from the threshold up to $W=2$ GeV.

\begin{figure}
\includegraphics[width=0.95\textwidth,clip]{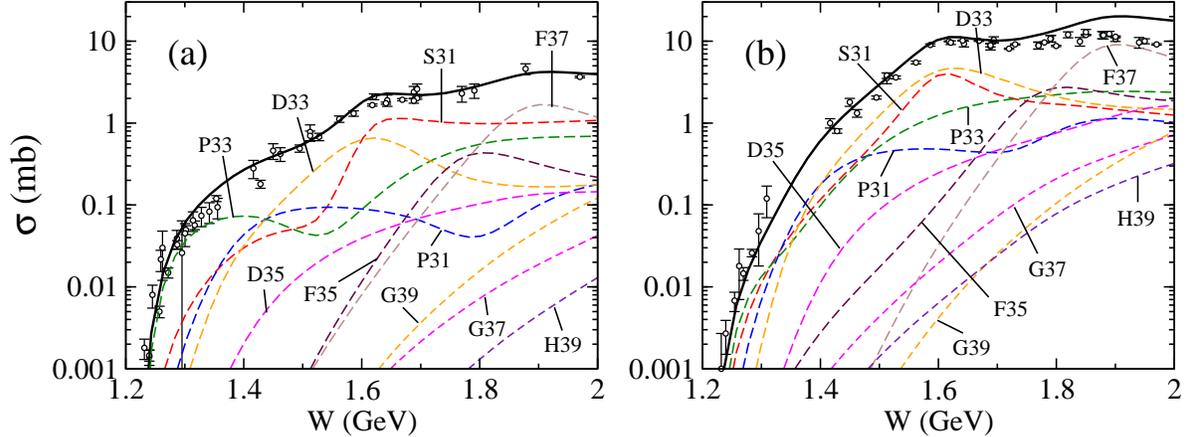}
\caption{
(Color online)
Contribution of each partial wave to the $\pi^+ p \to \pi \pi N$ total cross sections:
(a) $\pi^+ p \to \pi^+ \pi^+ n$ and
(b) $\pi^+ p \to \pi^+ \pi^0 p$.
The total cross section and the individual partial wave contributions
are shown as thick-solid and dashed curves, respectively.
See Refs.~\cite{manley,kjlms09} and references therein for the data.
\label{fig:pp-pwa-decomp}
}
\end{figure}
\begin{figure}
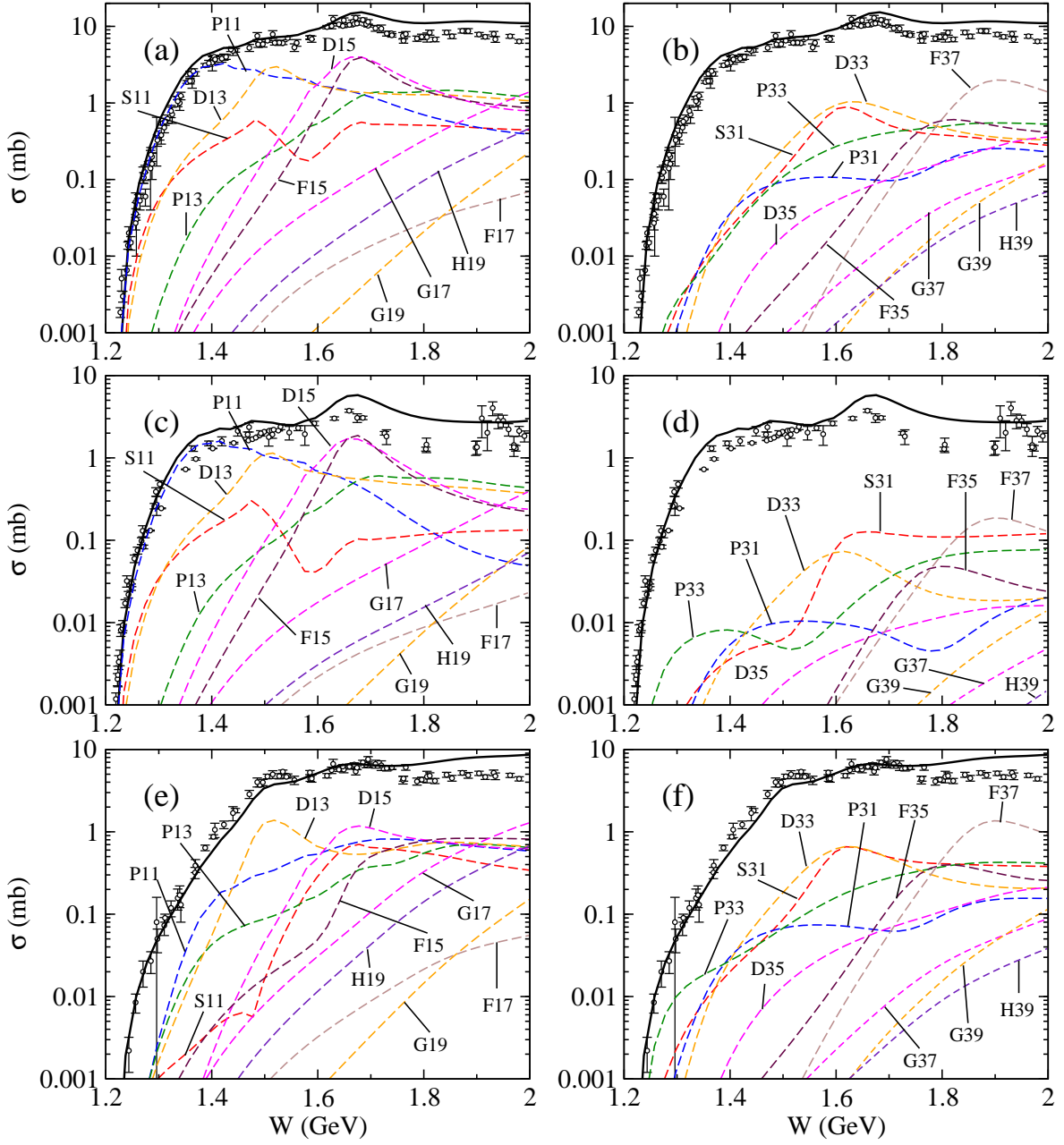

\includegraphics[width=0.95\textwidth,clip]{fig4-1}\\
\includegraphics[width=0.95\textwidth,clip]{fig4-2}\\
\includegraphics[width=0.95\textwidth,clip]{fig4-3}
\caption{
(Color online)
Contribution of each partial wave to the $\pi^- p \to \pi \pi N$ total cross sections.
Panels (a) and (b) $\pi^- p \to \pi^+ \pi^- n$;
(c) and (d) $\pi^- p \to \pi^0 \pi^0 n$;
(e) and (f) $\pi^- p \to \pi^- \pi^0 p$.
Panels (a), (c), and (e) [(b), (d), and (f)] present $I=1/2$ [$I=3/2$] partial waves for each reaction.
The total cross section and the individual partial wave contributions
are shown as thick-solid and dashed curves, respectively.
See Refs.~\cite{manley,kjlms09} and references therein for the data.
\label{fig:mp-pwa-decomp}
}
\end{figure}

In Figs.~\ref{fig:pp-pwa-decomp} and~\ref{fig:mp-pwa-decomp},
we present the contribution of each partial wave to the 
$\pi N\to\pi \pi N$ total cross sections.
The results for partial waves up to $J=9/2$ are plotted.
(Note that $\pi^+ p \to \pi \pi N$ contains only the isospin $I=3/2$ partial waves, 
while $\pi^- p \to \pi \pi N$ contains both the $I=1/2$ and $3/2$ partial waves.)
As for the initial $\pi^+ p$ reactions,
the $P_{33}$ partial wave dominates 
the $\pi^+ p \to \pi^+ \pi^+ n$ cross section
from the threshold up to  $W=1.4$~GeV, while 
a couple of partial waves equally contribute to 
$\pi^+ p \to \pi^+ \pi^0 n$ at low energies.
However, the $S_{31}$ and $D_{33}$ partial waves dominate the cross sections
of both the reactions in the $W=1.5$-$1.75$~GeV region;
above $W=1.8$~GeV the $F_{37}$ partial wave becomes dominant instead.
As for the initial $\pi^- p$ reactions, 
almost all the $I=3/2$ partial waves have only sub-dominant contributions
except for $\pi^- p \to \pi^+\pi^-n$ and $\pi^- p \to \pi^-\pi^0p$ at $W\sim 1.9$ GeV
where the $F_{37}$ can be comparable to the largest $I=1/2$ partial waves.
In the $W$ region between $1.5$ and $1.75$~GeV,
$D_{13}$ and $D_{15}$ are major partial waves for all the three charged states
of the initial $\pi^- p$ reactions, as is $F_{15}$
for the $\pi^+\pi^- n$ and $\pi^0\pi^0 n$ final states.
It should be emphasized that the $P_{11}$ partial wave dominates 
the cross sections of $\pi^- p \to \pi^+\pi^- n$ and $\pi^- p \to \pi^0\pi^0 n$ 
up to $W\sim 1.4$ GeV, where the Roper resonance exists.
This is in contrast to the photoproduction reactions,
for which the contribution of the Roper resonance 
is known to be minor on the cross sections
and is obscured by the significant contribution of the first $D_{13}$ resonance.
The $\pi N \to \pi \pi N$ reactions will thus provide crucial information
not only on the high-mass $N^\ast$ states, but also on the mysterious Roper resonance.
The importance of the $\pi N \to \pi \pi N$ data for the Roper resonance 
and the $P_{11}$ partial wave at low energies has also been discussed 
with various theoretical and phenomenological 
approaches (see, e.g., Refs.~\cite{roper1,roper2,roper3,roper4,roper5,roper6}).

\section{Examining potential impact of the $\pi N \to \pi \pi N$ data on 
determining the $N^\ast$ parameters}
\label{sec:f37}

Now we demonstrate a potential impact of the $\pi N\to \pi \pi N$ data 
on determining the $N^\ast$ resonance parameters.
For this purpose, we take the $F_{37}$ partial wave as an example.
As shown in Fig~\ref{fig:pp-pwa-decomp}, the contribution of this partial wave 
is dominant for $\pi^+ p \to \pi^+ \pi^+ n$ and $\pi^+ p \to \pi^+ \pi^0 p$ above $W=1.8$ GeV,
which is the energy region relevant to the J-PARC E45~\cite{hicks-prop} experiment.
Our procedure is as follows:
\begin{enumerate}
\item 
Construct a slightly different model from that developed in Ref.~\cite{dcc8}. 
For this purpose, we recall that in the DCC model of Ref.~\cite{dcc8}, 
each bare $N^\ast$ state has the following model parameters: 
the bare $N^\ast$ mass ($M_{N^*}^0$), 
the cutoffs for strong and electromagnetic interactions 
($\Lambda_{N^\ast}$ and $\Lambda_{N^\ast}^{\text{e.m.}}$),
the coupling constants for bare $N^\ast \to MB$ decays ($C_{{MB}(LS),N^\ast}$
where the $MB$ channel has the relative angular momentum $L$ and the total spin $S$),
and the bare $\gamma N \to N^\ast$ transition helicity amplitudes 
($\tilde A_{1/2}^{N^\ast}$ and $\tilde A_{3/2}^{N^\ast}$).
We first set 
the ``bare'' coupling constants of $N^\ast \to \pi\Delta$ of the $F_{37}$ partial wave 
to zero, i.e., $C_{{MB}(LS),N^\ast(F37)} = 0$ for all $LS$ states.
We then refit the $\pi N \to \pi N, K\Sigma$ and $\gamma N \to \pi N, K\Sigma$ data
by varying only the other parameters associated with the $F_{37}$ bare $N^\ast$ states,
i.e., $M_{N^*(F37)}^0$, $\Lambda_{N^\ast(F37)}$, $\Lambda_{N^\ast(F37)}^{\text{e.m.}}$,
$C_{MB(LS),N^\ast(F37)}$ with $MB=\pi N, \rho N, K\Sigma$,
$\tilde A_{1/2}^{N^\ast}$, and $\tilde A_{3/2}^{N^\ast}$, while
all of the remaining model parameters are kept fixed as those obtained in Ref.~\cite{dcc8}.
(Note that $F_{37}$ partial wave does not affect $\eta N$ and $K\Lambda$
production reactions with total isospin $I=1/2$ only.)
Hereafter we call the DCC model of Ref.~\cite{dcc8} the ``original DCC'', while
the refitted one the ``refitted DCC.''
\item 
Examine how the difference between the original and refitted DCC 
emerges in the $\pi N \to \pi \pi N$ observables.
\end{enumerate}
With this exercise, we can examine whether
the $\pi N \to \pi \pi N$ data provide crucial constraints
on the existing reaction models and 
the $N^\ast$ resonance parameters that are hard to determine
with the $\pi N, \gamma N \to \pi N, \eta N, K\Lambda, K\Sigma$ data only.

\begin{figure}
\centering
\includegraphics[width=0.8\textwidth,clip]{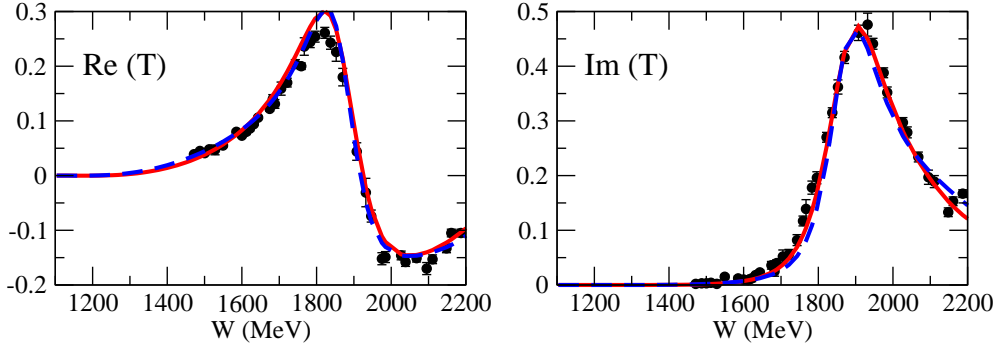}
\caption{
(Color online)
$F_{37}$ $\pi N$ partial wave amplitude.
Solid (red) curves are from the original DCC model~\cite{dcc8};
dashed (blue) curves show the refitted DCC model.
The data points are from the SAID energy-independent solution~\cite{said}.
\label{fig:f37}
}
\end{figure}
In Fig.~\ref{fig:f37}, we compare the $F_{37}$ $\pi N$ partial wave amplitudes
of the original and refitted DCC models.
One can see that both models reproduce the $F_{37}$ amplitudes with almost the same quality.
We have also confirmed that these models give almost the same $\chi^2$ values
for the $\pi N, \gamma N \to \pi N, K\Sigma$ data up to $W=2$ GeV.
This result indicates that the original and refitted DCC models
are hard to be distinguished from comparisons with the available
data for $\pi N, \gamma N \to \pi N, K\Sigma$ up to $W=2$ GeV.

For later use, we introduce the ``branching ratio'' $B_{MB}$ given by
\begin{equation}
B_{MB} = \frac{\gamma_{MB}}{\sum_{MB}\gamma_{MB}}.
\label{eq:br}
\end{equation}
Here, the ``partial decay width'' $\gamma_{MB}$ is defined for the stable meson-baryon channels 
($MB=\pi N, \eta N, K\Lambda, K\Sigma$) as
\begin{equation}
\gamma_{MB} = \rho_{MB}(\bar k;\bar M) \left| \bar\Gamma^R_{MB}(\bar k; \bar M)\right|^2,
\label{eq:gamma-stable}
\end{equation}
where 
$\rho (k;W) = \pi k E_M(k)E_B(k)/W$ with $E_\alpha(k) \equiv\sqrt{m_\alpha^2+k^2}$;
$\bar M = \mathrm{Re}(M_R)$ with $M_R$ being the complex pole mass of the $N^*$; and
$\bar k$ is given by $\bar M = E_M(\bar k) + E_B(\bar k)$.
The explicit expression of the dressed $N^*\to MB$ decay vertex
$\bar\Gamma^R_{MB}(k;W)$ has been given in Ref.~\cite{dcc8} and thus will not be presented here.
For the quasi-two-body channels of $\pi \pi N$ ($MB= \pi\Delta,\rho N,\sigma N$),
however, the $\gamma_{MB}$ is given as
\begin{equation}
\gamma_{\pi\Delta} = 
\frac{1}{2\pi}\int^{\bar M-m_\pi}_{m_\pi + m_N} dM_{\pi N}
\frac{-2\mathrm{Im}\bm{(}\Sigma_{\pi\Delta}(\bar k;\bar M)\bm{)}}
     {\left|\bar M - E_\pi(\bar k) - E_\Delta(\bar k)-\Sigma_{\pi\Delta}(\bar k;\bar M)\right|^2}
 \rho_{\pi\Delta}(\bar k;\bar M) \left| \bar\Gamma^R_{\pi\Delta}(\bar k; \bar M)\right|^2,
\label{eq:gamma-quasi}
\end{equation}
for the case of $MB = \pi \Delta$.
Here $\Sigma_{\pi\Delta}(k;W)$ is the self-energy in the $\pi\Delta$ Green's function
given in Ref.~\cite{dcc8}; $\bar k$ is defined by
$\bar M = E_\pi(\bar k) + \sqrt{M_{\pi N}^2 + \bar k^2}$ for the quasi-two-body channels.
The integral in Eq.~(\ref{eq:gamma-quasi}) accounts for the phase space of the final
$\pi \pi N$ states; Eq.~(\ref{eq:gamma-quasi}) reduces to Eq.~(\ref{eq:gamma-stable})
in the stable $\Delta$ limit: $\Sigma_{\pi\Delta}\to 0$.
A similar expression is obtained also for $MB = \rho N, \sigma N$.
The branching ratio defined in Eq.~(\ref{eq:br}) is a good measure of relative strength
of the coupling of a $N^*$ resonance to a meson-baryon channel $MB$
for clear resonances such as the first $F_{37}$ resonance as shown in Fig.~\ref{fig:f37}.

\begin{table*}[t]
\caption{\label{tab:f37}
Comparison of pole mass ($M_R$) and branching ratios ($B_{MB}$) 
for the decay to a channel $MB =\pi N, \pi\Delta, \rho N, K\Sigma$ 
of the $F_{37}$ nucleon resonance.
Allowed spin ($S$) and angular momentum ($L$) states for a given channel $MB$
are listed as ($L,S$) in the second row.
}
\begin{ruledtabular}
\begin{tabular}{lcccccccc}
&$M_R$ (MeV) & $B_{\pi N}$ (\%) & 
\multicolumn{2}{c}{$B_{\pi \Delta}$ (\%)} & 
\multicolumn{3}{c}{$B_{\rho N}$ (\%)} & $B_{K\Sigma}$ (\%) \\
 & &$(3,\frac{1}{2})$ &$(3,\frac{3}{2})$&$(5,\frac{3}{2})$&$(3,\frac{1}{2})$&$(3,\frac{3}{2})$&$(5,\frac{3}{2})$&$(3,\frac{1}{2})$\\
\hline
Original DCC~\cite{dcc8} &$1872-i103$ & $51.5$ & $46.7$ &$0.4$ &$1.1$&$0.1$&$0.1$&$0.0$\\
Refitted DCC         &$1867-i85$ & $53.5$ & $0.2$ &$0.0$ &$10.3$&$34.2$&$1.7$&$0.1$
\end{tabular}
\end{ruledtabular}
\end{table*}
The resulting resonance parameters of the first $F_{37}$ resonance 
are listed in Table~\ref{tab:f37}.
We find that the original and refitted DCC models give
just a slightly different pole mass: $|M_R^{\text{orig.}} - M_R^{\text{refit}}|\sim 19$ MeV.
However, the branching ratios for 
the decay to each component of the $\pi\pi N$ channel ($\pi\Delta$ and $\rho N$) 
are significantly different between the two models,
while the sum $B_{R,\pi\Delta} + B_{R,\rho N}$ is almost the same.
This suggests that a significant ambiguity may exist in their decay dynamics
even for clear resonances,
of which the pole mass is well determined by various analysis groups,
as far as the resonance parameters are extracted from the fit without 
including the double pion production data.

\begin{figure}[t]
\centering
\includegraphics[width=0.7\textwidth,clip]{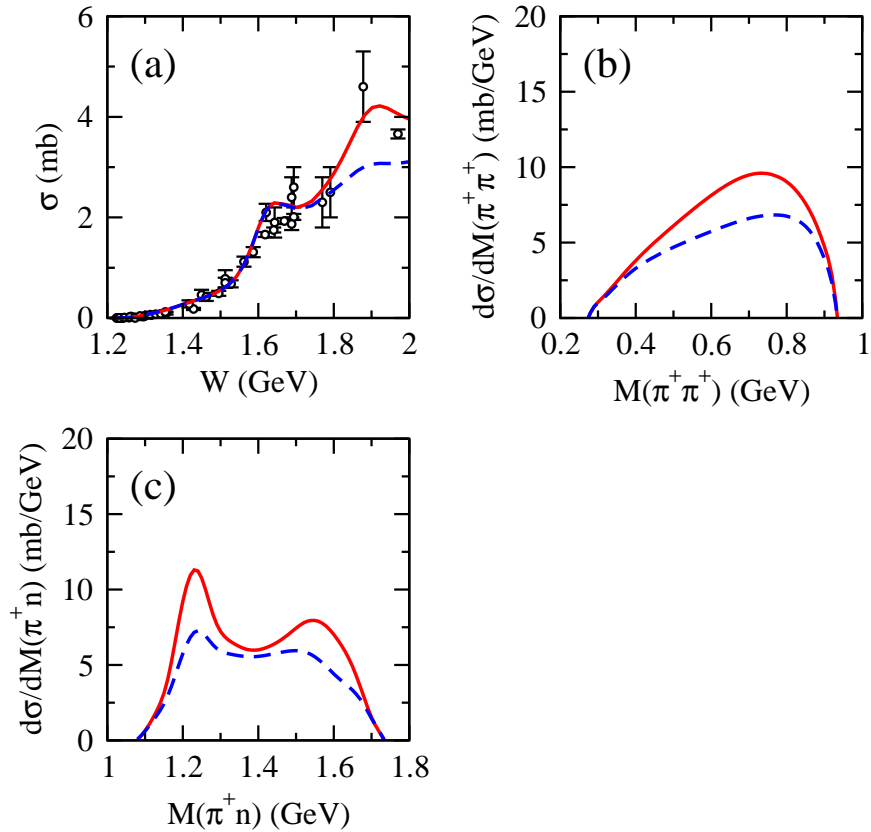}
\caption{
(Color online)
Comparison of the $\pi^+ p \to \pi^+\pi^+ n$ cross sections:
(a) total cross section; 
the invariant mass distributions of (b) $\pi^+ \pi^+$ and (c) $\pi^+ n$
at $W=1.87$ GeV.
Solid (red) curves are the original DCC; 
dashed (blue) curves are the refitted DCC.
See Refs.~\cite{manley,kjlms09} and references therein for the total cross section data.
\label{fig:ppn}
}
\end{figure}
\begin{figure}[t]
\centering
\includegraphics[width=0.7\textwidth,clip]{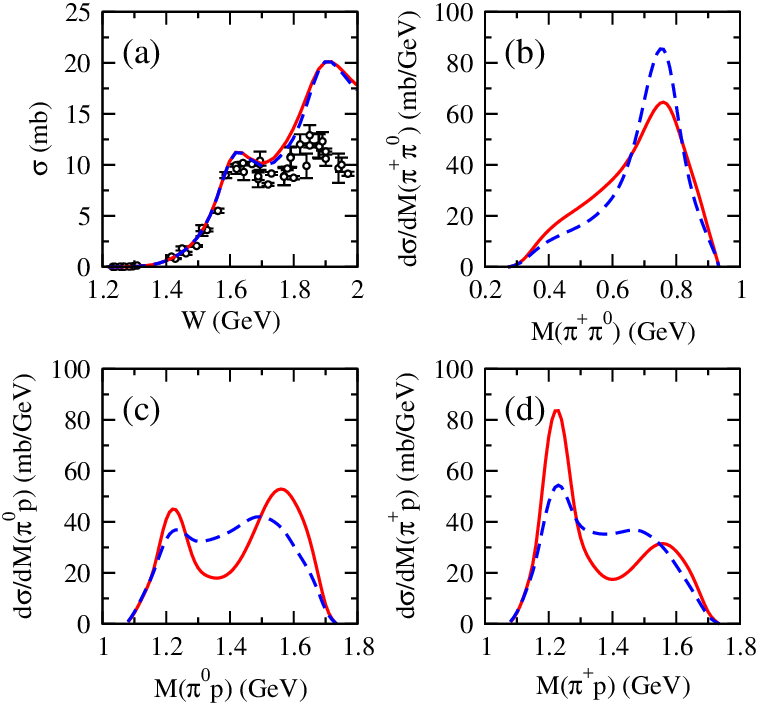}
\caption{
(Color online)
Comparison of the $\pi^+ p \to \pi^+\pi^0 p$ cross sections:
(a) total cross section; 
the invariant mass distributions of
(b) $\pi^+ \pi^0$, 
(c) $\pi^0 p$, and 
(d) $\pi^+ p$ at $W=1.87$ GeV.
Solid (red) curves are the original DCC; 
dashed (blue) curves are the refitted DCC.
See Refs.~\cite{manley,kjlms09} and references therein for the total cross section data.
\label{fig:00p}
}
\end{figure}
In Figs.~\ref{fig:ppn} and~\ref{fig:00p}, we present the total cross sections and 
invariant mass distributions of $\pi^+ p \to \pi^+\pi^+ n$ and $\pi^+ p \to \pi^+\pi^0 p$ 
calculated with both the original and refitted DCC models.
It is found that the two models show clear differences in those observables.
As for the $\pi^+ p \to \pi^+\pi^+ n$ reaction (Fig.~\ref{fig:ppn}),
the energy dependence of the total cross section is obviously different
above $W=1.7$ GeV, where the $F_{37}$ $N^\ast$ resonance exists.
As a result, the magnitude as well as the shape of the invariant mass distributions 
are also quite different between the two models.
On the other hand, the two models give almost the same 
total cross sections for $\pi^+ p \to \pi^+\pi^0 p$ (Fig.~\ref{fig:00p}).
However, the invariant mass distributions exhibit quite different shape
between the two models, although the integration of the distributions gives the
same total cross section.
It is noticed that the peak or bump in the refitted DCC model
seen at $M_{\pi\pi} \sim 0.75$ GeV in Fig.~\ref{fig:00p}(b)
is more enhanced than the original DCC, while
the peak at $M_{\pi N} \sim 1.2$ GeV in Fig.~\ref{fig:ppn}(c) and Figs~\ref{fig:00p}(c) and (d)
are less enhanced.
This behavior is consistent with the differences in the branching ratios 
between the two models shown in Table~\ref{tab:f37}.
These differences in the predicted observables will be large enough for 
distinguishing the two models if
the high statistics data of $\pi N \to \pi \pi N$ at J-PARC are obtained.
Also, this result indicates 
that the total cross sections are not enough 
and the differential cross section data such as the invariant mass distributions
are highly desirable for a quantitative study of the $N^\ast$ spectroscopy.
We have confirmed that clear differences between the two models
are observed also in the shape of invariant mass distributions
of $\pi^- p \to \pi^+\pi^- n$ and $\pi^- p \to \pi^-\pi^0 p$ at $W=1.87$ GeV, 
but not in $\pi^- p \to \pi^0\pi^0 n$ 
because of the minor contribution of $F_{37}$ to this reaction
[see Fig.~\ref{fig:mp-pwa-decomp} (c) and (d)].

Finally, we close this section with a remark about other partial waves.
We have made a similar examination also for 
other dominant partial waves, i.e., $S_{31}$ and $D_{33}$ in the initial $\pi^+ p$ reactions
and $D_{13}$, $D_{15}$, and $F_{15}$ in the initial $\pi^- p$ reactions.
We then find that at least the $N^\ast$ resonance parameters of $D_{15}$ and $F_{15}$ 
show uncertainties that are similar to $F_{37}$ discussed above and could be resolved 
once the precise $\pi N \to \pi \pi N$ data are available.

\section{Summary and outlook}
\label{sec:sum}

Motivated by an experimental proposal for the measurement 
of the $\pi N \to \pi \pi N$ reactions at J-PARC, 
we have examined a potential impact of the $\pi N \to \pi \pi N$ 
data on determining the resonance parameters associated with 
high-mass $N^\ast$ resonances, by making use of the predicted $\pi N\to\pi\pi N$ 
cross sections from the ANL-Osaka DCC model recently developed in Ref.~\cite{dcc8}.
We have found that the partial wave analysis 
without including the double-pion production data
would leave a sizable ambiguity in the $N^\ast$ resonance parameters,
particularly in those associated with the three-body $\pi \pi N$ channel,
and that the $\pi N \to \pi \pi N$ data will play a crucial role
for resolving the ambiguity.
The results in this work clearly show that
the measurement of the $\pi N \to \pi \pi N$ reaction, such as planned at J-PARC~\cite{hicks-prop},
is desirable for disentangling the high-mass $N^\ast$ resonances.
Once the precise and extensive data of $\pi N \to \pi \pi N$ at $W > 1.6$ GeV 
are available from the proposed experiment at J-PARC, we hope to extend our combined 
analysis immediately and make a detailed and quantitative examination of the role of 
the new $\pi N \to \pi \pi N$ data for the $N^*$ spectroscopy.
This will be presented elsewhere.

\acknowledgments

The author would like to thank Dr. T.-S.~H.~Lee and Dr. A.~M.~Sandorfi for careful reading 
of the manuscript and helpful comments, and Dr. T.~Sato for useful discussions.
This work was supported by JSPS KAKENHI Grant Number 25800149.
The author also acknowledges the support of the HPCI Strategic Program 
(Field 5 ``The Origin of Matter and the Universe'') of 
Ministry of Education, Culture, Sports, Science and Technology (MEXT) of Japan.
This work used resources of the
National Energy Research Scientific Computing Center, which is supported by the Office of
Science of the U.S. Department of Energy under Contract No. DE-AC02-05CH11231, and
resources provided on ``Fusion'', a 320-node computing cluster operated by the Laboratory
Computing Resource Center at Argonne National Laboratory.


\begin{thebibliography}{99}

\bibitem{manley}
D.~M.~Manley, R.~A.~Arndt, Y.~Goradia, and~V.L.~Teplitz,
Phys. Rev. D {\bf 30}, 904 (1984).

\bibitem{kjlms09}
H.~Kamano, B.~Juli\'a-D\'{\i}az, T.-S.~H.~Lee, A.~Matsuyama, and T.~Sato, 
Phys. Rev. C {\bf 79}, 025206 (2009).

\bibitem{add-exp1}
M. Kermani {\it et al.} (The CHAOS Collaboration),
Phys. Rev. C {\bf 58}, 3419 (1998).

\bibitem{add-exp2}
I.~G.~Alekseev {\it et al.},
Nucl. Phys. {\bf B541}, 3 (1999).

\bibitem{hicks-prop}
K.~Hicks and H.~Sako {\it et al.},
Measurements of $\pi N\to\pi\pi N$ and $\pi N \to KY$ at J-PARC
(J-PARC E45), http://j-parc.jp/researcher/Hadron/en/pac\_1207/pdf/P45\_2012-3.pdf.

\bibitem{neutrino}
H.~Kamano, S.~X.~Nakamura, T.-S.~H.~Lee, and T.~Sato,
Phys. Rev. D {\bf 86}, 097503 (2012).

\bibitem{neuint}
S.~X.~Nakamura, Y.~Hayato, M.~Hirai, H.~Kamano, S.~Kumano, M.~Sakuda, K.~Saito, and T.~Sato,
arXiv:1303.6032.

\bibitem{dcc8}
H.~Kamano, S.~X.~Nakamura, T.-S.~H.~Lee, and T.~Sato,
Phys. Rev. C {\bf 88}, 035209 (2013).

\bibitem{msl}
A.~Matsuyama, T.~Sato, and T.-S.~H.~Lee, 
Phys. Rep. {\bf 439}, 193 (2007).

\bibitem{jlms07}
B.~Juli\'a-D\'iaz, T.-S.~H.~Lee, A.~Matsuyama, and T.~Sato, 
Phys. Rev. C {\bf 76}, 065201 (2007).

\bibitem{roper1}
A.~V.~Sarantsev {\it et al.} (CB-ELSA and A2-TAPS Collaborations),
Phys. Lett. B {\bf 659}, 94 (2008).

\bibitem{roper2}
V.~Kozhevnikov and S.~Sherman,
Phys. Atom. Nucl. {\bf 71}, 1860 (2008).

\bibitem{roper3}
H.~Kamano and M.~Arima,
Phys. Rev. C {\bf 73}, 055203 (2006).

\bibitem{roper4}
T.~S.~Jensen and A.~F.~Miranda,
Phys. Rev. C {\bf 55}, 1039 (1997).

\bibitem{roper5}
V. Bernard, N. Kaiser, and U.-G. Mei{\ss}ner,
Nucl. Phys. {\bf B457}, 147 (1995).

\bibitem{roper6}
E.~Oset and M.~J.~Vicente-Vacas, 
Nucl. Phys. {\bf A446}, 584 (1985).

\bibitem{said}
CNS Data Analysis Center (GWU), http://gwdac.phys.gwu.edu;
R.~Arndt, W.~J.~ Briscoe, I.~I.~Strakovsky, and R.~L.~Workman,
Phys. Rev. C {\bf 74} 045205 (2006).


\end{thebibliography}
\end{document}